\documentclass[
	aps,
	twocolumn,
	groupetaddress,
	longbibliography,
]{revtex4-1}
\usepackage[english]{babel}
\usepackage[utf8]{inputenc}
\usepackage{amsmath,amssymb,amsfonts}
\usepackage{color}
\usepackage{graphicx}
\usepackage{physics}

\newcommand{\Sys}{\mathcal{S}}
\newcommand{\Env}{\mathcal{E}}
\newcommand{\Int}{\mathcal{I}}

\usepackage[
	colorlinks=true,
	linkcolor=blue,
	urlcolor=blue,
	citecolor=blue
]{hyperref}

\begin{document}

\title{Highly efficient and indistinguishable single-photon sources via phonon-decoupled two-color excitation}

\author{Luca Vannucci}
\email{lucav@dtu.dk}
\affiliation{DTU Electro, Department of Electrical and Photonics Engineering, 2800 Kongens Lyngby, Denmark}

\author{Niels Gregersen}
\affiliation{DTU Electro, Department of Electrical and Photonics Engineering, 2800 Kongens Lyngby, Denmark}

\date{\today}

\begin{abstract}

Single-photon sources with near-unity efficiency and indistinguishability play a major role in the development of quantum technologies.
However, on-demand excitation of the emitter imposes substantial limitations to the source performance.
Here, we show that coherent two-color pumping allows for population inversion arbitrarily close to unity in bulk quantum dots thanks to a decoupling effect between the emitter and its phonon bath.
Driving a micropillar single-photon source with this scheme, we calculate very high photon emission into the cavity mode (0.95 photons per pulse) together with excellent indistinguishability (0.975) in a realistic configuration, thereby removing the limitations imposed by the excitation scheme on single-photon source engineering.

\end{abstract}

\maketitle

\section{Introduction}
\label{sec:intro}

Photonic quantum technologies \cite{OBrien2009, Wang2019_review} --- such as quantum computers, simulators, and networks --- rely on multi-photon interference to process information \cite{Pan2012_RMP, Wang2019_BS, Sparrow2018} and therefore on the availability of efficient sources of indistinguishable single photons \cite{Gregersen2017}.
For a single-photon source (SPS) with photon output $\mathcal N$ and degree of indistinguishability $\mathcal I$, the rate of successful $n$-photon interference scales as $(\mathcal N \cdot \mathcal I)^n$ \cite{Gaal2022}. Thus, for scalable quantum information processing, the source's figure of merit $\mathcal N \cdot \mathcal I$ must be increased as close as possible to 1.

The most successful SPS is currently based on cavity-coupled semiconductor quantum dots (QDs) \cite{Somaschi2016, Wang2019_source, Tomm2021},
which are however strongly affected by the vibrational environment.
Previous theoretical work demonstrates a trade-off between $\mathcal N$ and $\mathcal I$ induced by phonon scattering, and indicates a pathway towards optimal performance using the cavity effect \cite{Iles-Smith2017}. By carefully engineering the cavity, simulations predict values of $\mathcal N \cdot \mathcal I$ in the range 0.95--0.98 once the emitter is initialized in the excited state \cite{Wang2020_PRB_Biying, Gaal2022}.
This calls for an excitation scheme that prepares the desired initial state with the highest possible fidelity and is compatible with the requirement $\mathcal N \cdot \mathcal I \to 1$.
In this paper, we show that the performance of a SPS driven with two-color excitation schemes \cite{He2019, Koong2021, Bracht2021, Karli2022} matches the one calculated for an initially excited source, thereby demonstrating that the excitation scheme is no longer a limitation.

Initial experiments on SPSs relied on $p$-shell pumping \cite{Gazzano2013, Somaschi2016}, whereby a laser pulse excites the QD into a higher energy state, which subsequently decays to the exciton level. Owing to the shorter wavelength of the pump with respect to the outgoing single photons, the laser is then removed via spectral filtering.
However, indistinguishability obtained under $p$-shell excitation is significantly reduced by the time-jitter effect \cite{Kiraz2004}.
Electrical triggering, which has been explored as an alternative to optical pumping \cite{Patel2010, Schlehahn2016}, suffers from a similar mechanism \cite{Kaer2013}.
Resonant excitation with short laser pulses set a new milestone, enabling two-photon interference visibility $\geq 0.99$ \cite{Somaschi2016, Ding2016}.
A resonant scheme, however, requires cross-polarization filtering to distinguish the outgoing single photons from the pump. This, in turn, suppresses the number of available photons by at least a factor 2, so that $\mathcal N \cdot \mathcal I$ can never exceed 0.5.

A trade-off between these two competing effects is offered by near-resonant phonon-assisted excitation, where $\mathcal N = 0.50$ has been demonstrated in experiments at the expense of a lower $\mathcal I = 0.91$ \cite{Thomas2021}.
Still, exciton preparation is limited to 0.85--0.90 fidelity both in experiments and theory \cite{Thomas2021, Cosacchi2019, Gustin2020}, posing a fundamental limitation towards further increasing the performance.
A promising strategy involving stimulated emission from the biexciton level \cite{Sbresny2022, Wei2022_stimulatedTPE, Yan2022_stimulatedTPE} has generated single photons with $\mathcal I = 0.93$ and in-fiber efficiency of 0.51 \cite{Wei2022_stimulatedTPE} in experiments.
However, increasing the figures of merit towards unity demands a detailed understanding of the role of phonons during the excitation process \cite{Luker2019, Bracht2022_pssb}, and an excitation scheme which is compatible with arbitrary increase towards unity of $\mathcal N \cdot \mathcal I$ has not been demonstrated so far.

\begin{figure}
	\centering
	\includegraphics[width=0.9\linewidth]{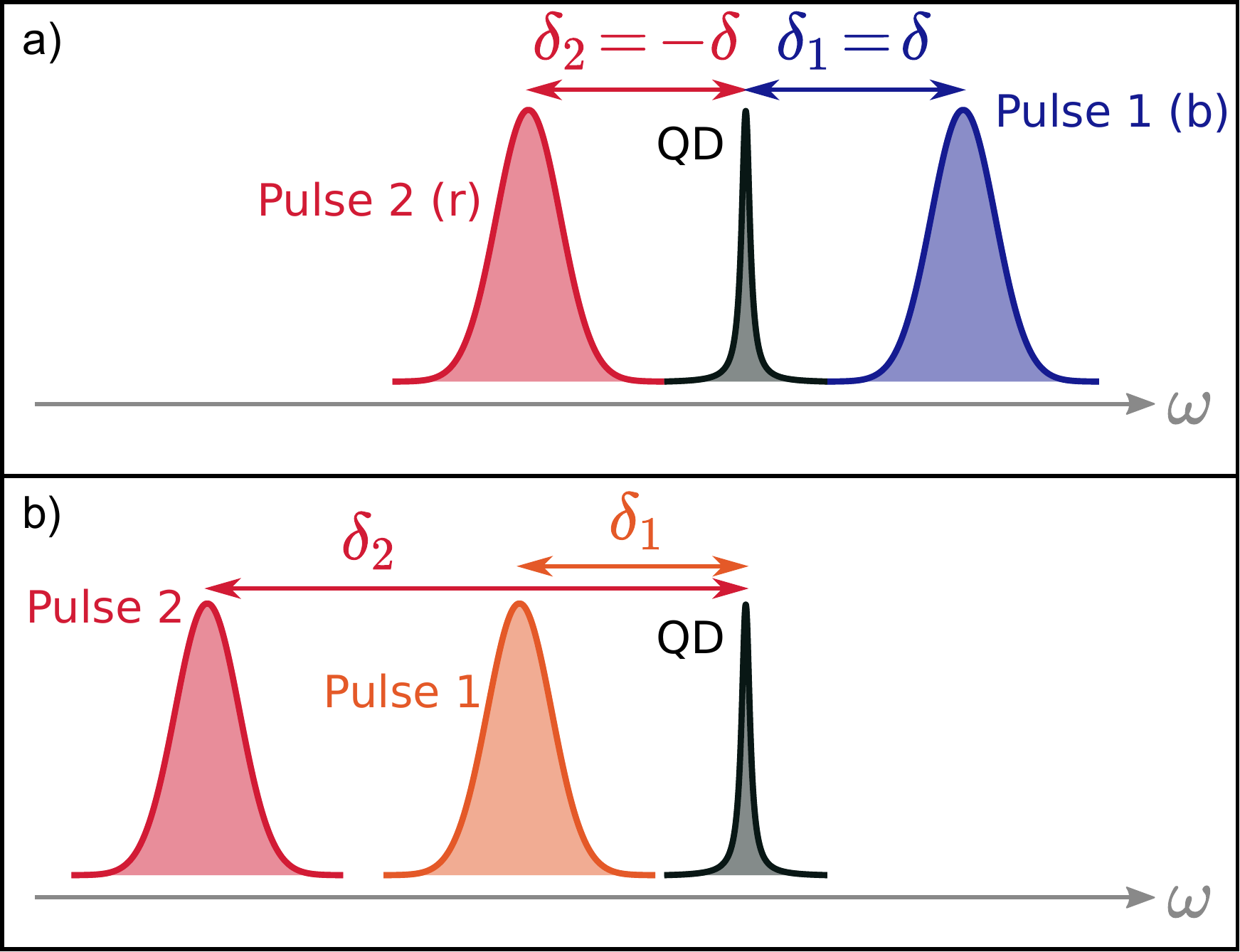}
	\caption{
		Sketch of two-color excitation schemes in the frequency domain relative to the QD emission line. a) ``Red-and-blue'' dichromatic excitation. b) SUPER scheme.
	}
	\label{fig:1}
\end{figure}

A dichromatic (or two-color) protocol makes use of two laser pulses detuned from the QD emission frequency (Fig.\ \ref{fig:1}a).
He \textit{et al.} initially proposed to use a symmetric ``red-and-blue'' configuration --- that is, with one pulse on the blue side and one on the red side of the spectrum with respect to the emitter \cite{He2019} --- and Koong \textit{et al.}\ subsequently demonstrated partial population of the exciton level by acting on the relative pulse amplitudes \cite{Koong2021}.
This effect is however significantly hindered by phonon scattering, with a population inversion $\sim$0.6 predicted theoretically in Ref.\ \cite{Koong2021}.
Here, on the other hand, we access a phonon-decoupled regime where phonon effects are removed from the excitation process \cite{Vagov2007, Kaldewey2017} and arbitrarily high population inversion is within reach for bulk QDs.
As a specific example, we demonstrate that a micropillar SPS driven with our scheme can generate up to 0.95 photons per pulse into the collection optics with even better indistinguishability than the one obtained under resonant excitation.

An alternative two-color strategy named SUPER scheme \cite{Bracht2021, Karli2022, Boos2022, Bracht2023, Heinisch2023} makes use of two laser pulses on the red side of the spectrum (Fig.\ \ref{fig:1}b).
This has resulted in an estimated population inversion ranging from 0.67 \cite{Boos2022} to 0.97 \cite{Karli2022} in experiments, but insufficient indistinguishability to date \cite{Boos2022}.
As we show in the following, the SUPER scheme is also compatible with an increase of $\mathcal N \cdot \mathcal I$ towards 1, provided that the phonon decoupled regime is attained \cite{Bracht2022_pssb}.

The paper is organized as follows. In Section \ref{sec:bulk} we study the population inversion of a bulk quantum dot under two-color excitation. We consider both the ``red-and-blue'' and the SUPER scheme, and we show the influence of phonon coupling on the performance of both schemes. Then, in Section \ref{sec:SPS} we calculate the photon output and the indistinguishability of a state-of-the-art SPS driven with two-color excitation, and we compare with the artificial case of an initially excited emitter. In Section \ref{sec:conclusions} we draw our conclusions. 
Four Appendices are devoted to technical details and to the validation of our methods.

\section{Population inversion of a bulk quantum dot}
\label{sec:bulk}

\subsection{Red-and-blue dichromatic scheme}

We begin by considering the dichromatic pumping dynamics of a QD in bulk in the absence of phonon coupling.
We thus take a two-level system --- ground state $\ket{G}$, excited state $\ket{X}$ --- which is coupled to two laser pulses centered at angular frequency $\omega_j$, $j \in \{1,2\}$.
They have Gaussian shape in the time domain, namely
\begin{equation}
	\Omega_j(t) = \frac{\Theta_j}{t_p \sqrt{\pi}} e^{-\qty(\frac{t}{t_p})^2} ,
\end{equation}
where $\Theta_j = \int_{-\infty}^{+\infty} \dd{t} \Omega_j(t)$ is the pulse area, and $t_p$ is the pulse temporal width (identical for both pulses, for simplicity).
In a reference frame rotating at the exciton frequency $\omega_X$, and making use of the rotating wave approximation, the system Hamiltonian reads
\begin{equation}
\label{eq:H_no_ph}
	H_\Sys(t) = \frac{\hbar}{2} \qty[ \Omega_1(t) e^{-i \delta_1 t} + \Omega_2(t) e^{-i \delta_2 t}] \sigma^\dag + \mathrm{h.c.},
\end{equation}
where $\sigma^\dag = \dyad{X}{G}$ is the QD raising operator, and $\delta_j = \omega_j - \omega_X$ is the frequency detuning from the exciton.
The time evolution of the density operator $\rho_\Sys (t)$ is readily obtained by solving the Von Neumann equation $\dot \rho_\Sys(t) = - \frac{i}{\hbar} \comm{H_\Sys(t)}{\rho_\Sys(t)}$, with the QD initialized in the ground state at $t=t_0$.
For the moment, we neglect the QD spontaneous emission and any source of decoherence to illustrate the physics of the pumping mechanism.

\begin{figure}
	\centering
	\includegraphics[width=0.99\linewidth]{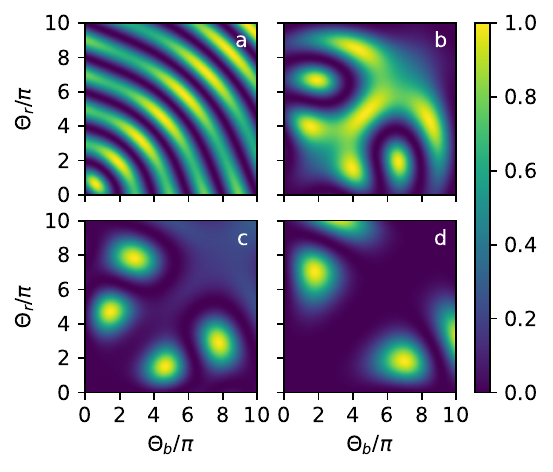}
	\caption{
		Exciton population in the absence of phonon coupling after the pulse as a function of $\Theta_b$ and $\Theta_r$, and for $\eta = 1$ (a), $\eta = 3$ (b), $\eta = 4$ (c), $\eta = 6$ (d).
	}
	\label{fig:2}
\end{figure}

First, let us focus on the red-and-blue configuration discussed in Refs.\ \cite{He2019, Koong2021} with symmetric detuning $\delta_1 = -\delta_2 = \delta$.
In the following discussion, we will use the notation $1 \equiv b$ and $2 \equiv r$ to identify the first and second pulse with the blue and red side of the spectrum, respectively.
To assess the pumping efficiency, we consider as a figure of merit the exited state population $P_X(t) = \Tr \qty[\sigma^\dag \sigma \rho_\Sys(t)]$ at a time $t$ after the pulse is gone (specifically, we use $t=3t_p$).
In the ideal scenario where any source of decoherence and dissipation is neglected, $P_X$ can take a maximum value of $P_X=1$, corresponding to perfect population inversion.
When the system dynamics is unitary and governed by Eq.\ \eqref{eq:H_no_ph}, one can show that $P_X$ after the laser pulse is determined by $f(\Theta_b, \Theta_r, \eta = t_p \delta)$, i.e.\ it depends on the product $\eta = t_p \delta$ and not on $t_p$ and $\delta$ separately (see Appendix \ref{app:proof}).
We explore such a functional dependence in Fig.\ \ref{fig:2}.
At $\eta = 1$ (Fig.\ \ref{fig:2}a) we observe a periodic pattern that is reminiscent of Rabi oscillations, especially along the diagonal $\Theta_b = \Theta_r$. Indeed, the symmetric dichromatic driving with $\Omega_b(t) = \Omega_r(t) = \Omega(t)$ has a simple analytical solution $P_X = \sin^2(\xi)$, where $\xi = \int_{-\infty}^{+\infty} \dd{t} \Omega(t) \cos(\delta t)$ is the spectral component of the dichromatic laser pulse at the exciton frequency \cite{Koong2021}.
This shows that the oscillations along the diagonal are due to direct resonant coupling to the QD, which gives rise to the typical Rabi physics.

The spectral component of the driving laser at the exciton frequency, which scales as $e^{-\eta^2/4}$, is smaller at larger $\eta$. 
Here, richer physics is observed in the exciton population. Rabi oscillations along the diagonal become progressively slower --- one full oscillation is visible in Fig.\ \ref{fig:2}b, while almost no oscillations are observed in Figs.\ \ref{fig:2}c and \ref{fig:2}d. At the same time, new bright spots exhibiting $P_X = 1$ emerge at $\Theta_b \ne \Theta_r$, which are not due to direct resonant excitation \cite{Koong2021}.
Their distance from the origin increases with $\eta$ and is linked to the total power provided by the laser pulse, which scales as $\sim (\Theta_b^2 + \Theta_r^2) / t_p$.
Therefore, a trade-off between larger values of $\eta$ --- ensuring low spectral overlap of the laser with the exciton frequency --- and lower values to minimize the power is necessary.
We use $\eta = 6$ in the following, for which the laser spectral component at $\omega_X$ is $e^{-6^2/4} \approx 1 \cdot 10^{-4}$ relatively to its peak value.

We analyze now the performance of the dichromatic driving at $\eta=t_p \delta=6$ in the presence of phonon-induced dissipation, focusing on the case of GaAs as host material.
The QD couples to a phonon environment --- represented by $H_\Env = \sum_k \hbar \nu_k b^\dag_k b_k$ --- through the interaction Hamiltonian
\begin{equation}
	\label{eq:phonon_int}
	H_\Int = \sum_k \hbar g_k (b^\dag_k + b_k) \dyad{X} .
\end{equation}
The environment is characterized by a phonon spectral density $J_{\rm ph}(\omega) = \sum_k |g_k|^2 \delta(\omega - \nu_k) \approx \alpha \omega^3 e^{-\omega^2 / \omega_{\rm c}^2}$. For a QD in GaAs we use the coupling strength $\alpha = 0.03$\,ps$^2$, and the frequency cutoff $\omega_{\rm c} = 2.2$\,THz \cite{Nazir2016, Denning2020_OSA}.
The effect of the phonon bath is, among other things, to shift the exciton frequency to $\omega_X - D$, with (see e.g.\ Ref.\ \cite{Nazir2016})
\begin{equation}
	D = \int_0^{+\infty} \dd{\omega} \frac{J_{\rm ph}(\omega)}{\omega} = \frac{\sqrt{\pi}}{4} \alpha \omega_{\rm c}^3 .
\end{equation}
It is convenient to move into a frame rotating at frequency $\omega_X - D$, where the system Hamiltonian reads
\begin{multline}
	H_\Sys(t) = -\hbar D \dyad{X} \\
	+ \frac{\hbar}{2} \sum_{j=1,2} \qty[ \Omega_j(t) e^{-i \delta_j t} \sigma^\dag + \Omega_j(t) e^{+i \delta_j t} \sigma] ,
\end{multline}
with $\delta_j = \omega_j - \omega_X + D$ the frequency detuning of each pulse with respect to the re-normalized exciton frequency.
The dynamics in the absence of phonon coupling is readily obtained by setting $\alpha=0$ (and thus $D=0$).

\begin{figure}
	\centering
	\includegraphics[width=0.99\linewidth]{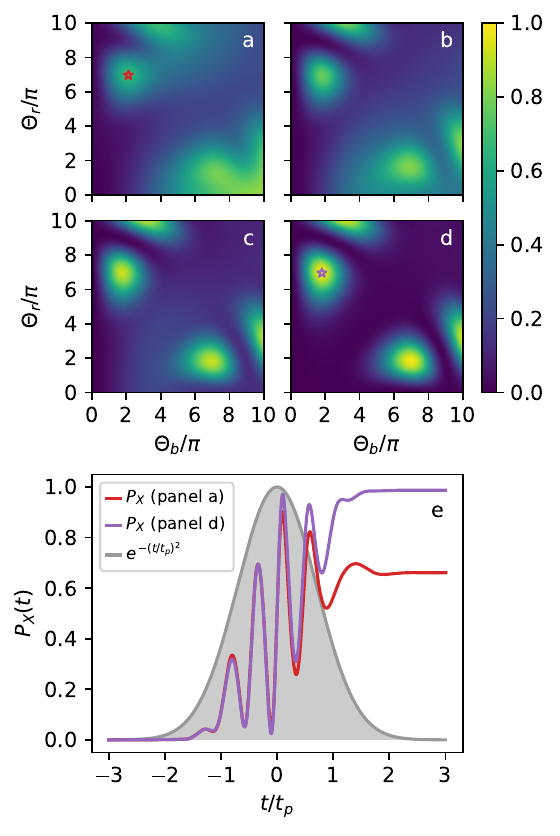}
	\caption{
		(a-d) Exciton population after the pulse as a function of $\Theta_b$ and $\Theta_r$, for fixed $\eta = t_p \delta = 6$ and $\delta = 1$\,THz (a), $\delta = 2$\,THz (b), $\delta = 3$\,THz (c), $\delta = 6$\,THz (d). Here, phonon-induced effects are considered.
		(e) $P_X(t)$ for $t_p = 6$\,ps, $\delta = 1$\,THz, $\Theta_b = 2.12\pi$, $\Theta_r = 6.96\pi$ (red), $t_p = 1$\,ps, $\delta = 6$\,THz, $\Theta_b = 1.80\pi$, $\Theta_r = 6.96\pi$ (purple).
	}
	\label{fig:3}
\end{figure}

To obtain $\rho_\Sys(t)$, we adopt a master equation formalism within the weak-coupling approximation \cite{Nazir2016}, whose implementation is detailed in Appendix \ref{app:methods}. The reduced density operator is determined by solving
\begin{equation}
	\dv{t} \rho_\Sys(t) = - \frac{i}{\hbar} \comm{H_\Sys(t)}{\rho_\Sys(t)} + \mathcal{K} \qty[\rho_\Sys(t)]
\end{equation}
where the extra term $\mathcal{K}$ accounts for environment-induced effects.
Due to the intrinsic asymmetry between phonon absorption and emission at low temperature, $P_X$ is now function of $t_p$ and $\delta$ separately.
The behavior of $P_X$ for $t_p = 6$\,ps and $\delta = 1$\,THz (corresponding to $\hbar \delta \approx 0.65\,\mathrm{meV}$) is reported in Fig.\ \ref{fig:3}a, where the features described in Fig.\ \ref{fig:2}d are significantly hindered by phonon scattering. For an excitation pulse predominantly on the blue side (i.e.\ $\Theta_b > \Theta_r$), we observe a rather broad area revealing $P_X \approx 0.8$, which is attributed to phonon-assisted processes \cite{Cosacchi2019, Gustin2020, Thomas2021}.
On the other hand, the red side ($\Theta_b < \Theta_r$) shows an isolated peak similarly to the case without phonon coupling, but with a significantly smaller maximum value. 
We find $P_X = 0.661$ at $(\Theta_b, \Theta_r) = (2.12\pi, 6.96\pi)$, which is in qualitative agreement with Ref.\ \cite{Koong2021}, where a similar detuning was used.

Moving to larger detuning (while simultaneously keeping $\eta = t_p \delta$ fixed), phonon scattering becomes progressively less detrimental.
In Figs.\ \ref{fig:3}b, \ref{fig:3}c, and \ref{fig:3}d, the dichromatic features gradually emerge from the background, with Fig.\ \ref{fig:3}d begin almost identical to the corresponding calculation in the absence of phonons (Fig.\ \ref{fig:2}d). Indeed, at $\delta = 6$\,THz ($\hbar \delta \approx 3.95\,\mathrm{meV}$) we find a maximum $P_X = 0.987$ at $(\Theta_b, \Theta_r) = (1.80\pi, 6.96\pi)$.
One can further increase $P_X$ arbitrarily close to 1 by increasing the detuning  beyond $\delta = 6$\,THz at constant $\eta = t_p \delta$. For instance, we find $P_X=0.999$ at $(t_p, \delta)$ = (0.2\,ps, 30\,THz) (see Appendix \ref{app:temperature}).
However, a configuration with such ultra-short pulses and large detuning has practical challenges, and we will limit the discussion to the range $t_p \in [1, 6]$\,ps.

We interpret this result in terms of an effective phonon decoupling occurring at larger $\delta$ and shorter $t_p$, which is known to play a role in the reappearance of Rabi oscillations at large pump power and in the Adiabatic Rapid Passage \cite{Vagov2007, Kaldewey2017, Reiter2019, Luker2019}. The population inversion is the result of a fast oscillating dynamics illustrated in Fig.\ \ref{fig:3}e, where we plot $P_X(t)$ for the two configurations marked with a star in Figs.\ \ref{fig:3}a and \ref{fig:3}d.
We observe that the excited state population oscillates on a time scale which is shorter than $t_p$.
However, phonon relaxation occurs on a time scale of $\sim$1--5\,ps \cite{Nazir2016, Barth2016}.
For $(t_p, \delta)$ = (6\,ps, 1\,THz) (red), the dynamics is sufficiently slow to allow for phonon-mediated relaxation events, and $P_X$ remains well below 1 after the dichromatic pulse.
For $(t_p, \delta)$ = (1\,ps, 6\,THz) (purple), on the other hand, such oscillations occur on a time scale that is much shorter than the phonon dynamics. Phonons cannot follow the QD dynamics instantaneously and are effectively decoupled from the emitter, resulting in very little dissipation effect and much higher population inversion (a phenomenon that also occurs for resonant excitation \cite{Reiter2014}) Temperature has almost no influence on the exciton preparation, further confirming the decoupling effect (see Appendix \ref{app:temperature}).
At the same time, the phonon spectral density is mostly contained within a range of 2--3\,THz. When moving to $\delta=6$\,THz, the detuning becomes larger than the maximum phonon frequency and no states are available for phonon emission or absorption. This explains why phonon-assisted events (which are particularly evident in the bottom right region of Fig.\ \ref{fig:3}a) are no longer allowed at larger detuning.
Note that the value $\delta=6$\,THz stays well below the $p$-shell of Stranski-Kranstanov grown dots. For instance, in Ref.\ \cite{Gazzano2013} the $p$-shell is $53.7$\,THz above the exciton level in our units, so that the blue pulse will not excite higher confined states.
Similarly, biexciton preparation via the two-photon resonance occurs roughly 1\,THz below the exciton line \cite{Wei2022_stimulatedTPE} and will not interfere with our optimal scheme.
It should be noted, however, that QD of different type and size may present a different phonon spectral density \cite{Luker2017} or different $p$-shell energy \cite{Lehner2023}.

The ability to capture the decoupling effect numerically is crucial.
We have used here the weak-coupling model due to its straightforward formulation and ease of implementation, however more advanced methods are available in the literature \cite{Strathearn2018, Cygorek2022}.
In Appendix \ref{app:comparison}, we have thus tested the weak-coupling model against the numerically exact TEMPO method \cite{Strathearn2018, OQuPy}, obtaining excellent agreement. The polaron master equation \cite{Iles-Smith2017, Denning2020_OSA}, on the other hand, overestimates the detrimental effect of phonons on the exciton preparation by roughly $5\%$ at short $t_p$.

\subsection{SUPER scheme}

\begin{figure}
	\centering
	\includegraphics[width=0.99\linewidth]{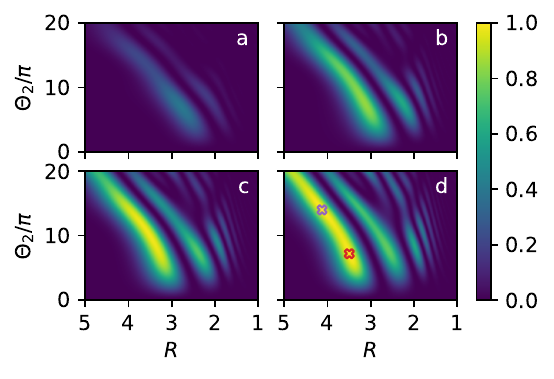}
	\caption{
		Exciton population obtained under the SUPER scheme as a function of $R = \delta_2 / \delta_1$ and $\Theta_2$, for fixed $\eta = t_p \delta_1 = -6$. We use $\Theta_1 = 4\pi$ (a), $\Theta_1 = 6\pi$ (b) $\Theta_1 = 7\pi$ (c), and $\Theta_1 =  8\pi$ (d). Here, the phonon coupling is not included in the calculation.
	}
	\label{fig:4}
\end{figure}

In the SUPER scheme, both laser pulses are spectrally positioned to the red side of the emitter (i.e. $\delta_2 < \delta_1 < 0$) \cite{Bracht2021}.
In the absence of phonons, the population inversion $P_X$ is determined by a function $f$ of four parameters $f(\Theta_1, \Theta_2, \eta, R)$, with $\eta = t_p \delta_1$ and $R=\delta_2/\delta_1$ (see Appendix \ref{app:proof}). As frequently done in the literature \cite{Bracht2021, Karli2022, Bracht2023, Boos2022}, one can fix the parameters pertaining to the first pulse (time duration, detuning, amplitude) and explore the functional dependence with respect to the second pulse.
We perform this analysis in Fig.\ \ref{fig:4}, where we fix the value $\eta = t_p \delta_1 = -6$ and use different values of the first-pulse amplitude $\Theta_1$ in each panel.
We observe that a threshold amplitude is required in order to reach full population inversion, i.e. we obtain $P_X=1$ only for $\Theta_1 \geq 8 \pi$ (panel d).
Two maxima are observed at values $\Theta_1 = 8\pi, \Theta_2 = 7.2 \pi, \eta=-6, R=3.495$ (red cross) and $\Theta_1 = 8\pi, \Theta_2 = 14.0 \pi, \eta=-6, R=4.125$ (purple cross). Up to this point, any combination of $t_p$, $\delta_1$ and $\delta_2$ generating the same values of $\eta$ and $R$ leads to perfect population inversion.

\begin{figure}
	\centering
	\includegraphics[width=0.99\linewidth]{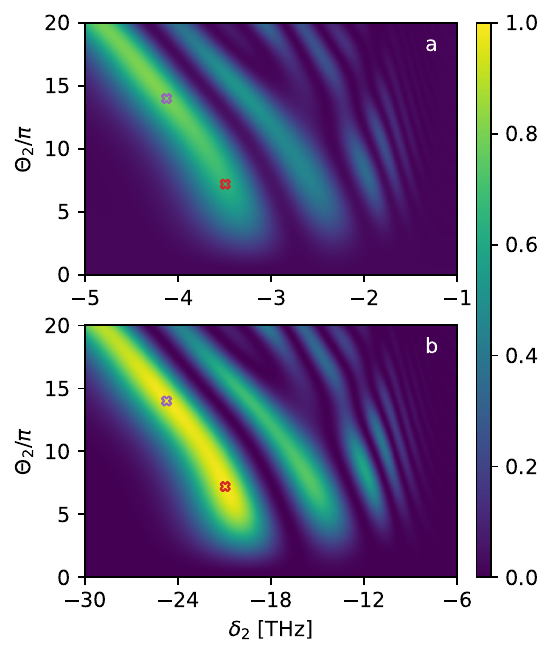}
	\caption{
		Exciton population obtained under the SUPER scheme as a function of $\delta_2$ and $\Theta_2$, for fixed $\eta = t_p \delta_1 = -6$ and $\Theta_1 = 8\pi$, and with the inclusion of phonon coupling.
		We use $t_p$= 6\,ps (a) and $t_p$= 1\,ps (b), with $\delta_1 = \eta / t_p$ determined accordingly.
	}
	\label{fig:5}
\end{figure}

However, this symmetry is broken by the phonon coupling. In Fig.\ \ref{fig:5} we compare the two cases $(t_p, \delta_1)$ = (6\,ps, $-$1\,THz) and $(t_p, \delta_1)$ = (1\,ps, $-$6\,THz) (panels a and b, respectively) in the presence of phonon coupling at fixed $\Theta_1=8\pi$. A significant quantitative difference is found.
For the same values of $\Theta_2$ and $R$, we now obtain $P_X$=0.653 (red cross) and 0.795 (purple cross) for longer pulses (panel a) compared to $P_X$=0.989 (red cross) and 0.985 (purple cross) for shorter pulses (panel b). This is explained again in terms of phonon decoupling.
The former configuration is relatively slow compared to the phonon relaxation time and uses a frequency detuning smaller than the phonon spectral density, while the latter uses sufficiently short pulses and large detuning to effectively avoid phonon scattering. 
As before, these figures can be pushed arbitrarily close to 1 by decreasing $t_p$ and increasing $\delta_1$ and $\delta_2$ by identical factors --- for instance, we obtain $P_X = 0.997$ at $t_p$= 0.5\,ps.

\begin{figure}
	\centering
	\includegraphics[width=0.99\linewidth]{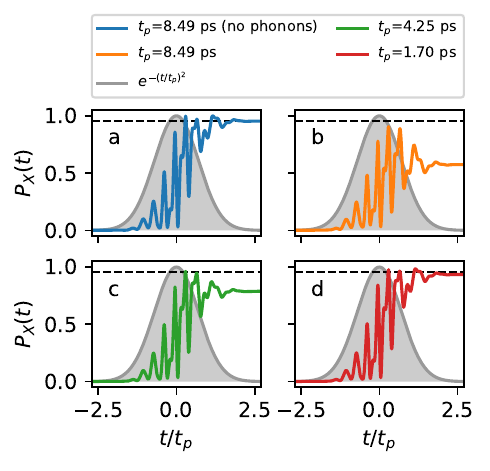}
	\caption{
		(a) Evolution of the exciton population in time for $\Theta_1 = 8 \pi$, $\Theta_2 = 8.8 \pi$, $t_p = 8.49$\,ps, $\delta_1=-1.064$\,THz, $\delta_2=-3.115$\,THz, and no phonon coupling.
		(b-d) Same calculation in the presence of phonon coupling, with $t_p = 8.49$\,ps (b), $t_p = 4.25$\,ps (c), $t_p = 4.25$\,ps (d). In (b-d) the values of $\delta_1$ and $\delta_2$ are adjusted to keep $\eta = t_p \delta_1 = -9.02$ and $R = \delta_2 / \delta_1 = 3.31$ fixed. The dashed line marks the performance in the absence of phonons.
	}
	\label{fig:6}
\end{figure}

We note that phonon scattering can possibly explain some observation reported by Boos et al.\ in Ref.\ \cite{Boos2022}. There, a population inversion of $P_X \approx 0.66$ is estimated experimentally for $\Theta_1 = 8 \pi$, $\Theta_2 = 8.8 \pi$, $t_p = 8.49$\,ps, $\delta_1=-1.064$\,THz, $\delta_2=-3.115$\,THz in our units (corresponding to $\eta = -9.02$ and $R = 3.31$), whereas a near-unity $P_X$ is expected in the absence of phonon coupling.
In Figs.\ \ref{fig:6}a and \ref{fig:6}b we report the dynamics both in the absence and in the presence of phonon coupling for such a configuration, obtaining final values of $P_X = 0.954$ and $P_X = 0.574$ respectively. This shows qualitatively that the imperfect population inversion reported in Ref.\ \cite{Boos2022} is partly due to phonon coupling.
Indeed, the pulse duration $t_p = 8.49$\,ps is rather long compared to the phonon relaxation time, and the pulse detunings $\delta_1$ and $\delta_2$ are well within the phonon spectral density.
Better performance could be obtained by reducing $t_p$ while correspondingly increasing $\delta_1$ and $\delta_2$. For instance, we obtain $P_X = 0.789$ by halving $t_p$ at constant $\eta$ and $R$ (Fig.\ \ref{fig:6}c), and $P_X = 0.935$ when $t_p$ is reduced by a factor 5 (Fig.\ \ref{fig:6}d).

\section{Single-photon source figures of merit: photon output and indistinguishability}
\label{sec:SPS}

We now characterize a state-of-the-art SPS driven with the two-color scheme in terms of photon output $\mathcal N$ and indistinguishability $\mathcal I$, and compare to the case of resonant pumping.
State-of-the-art SPSs rely on the cavity effect to funnel the emission into the zero-phonon line and direct the outgoing photons towards the collection optics. 
We thus introduce a single-mode cavity with annihilation operator $a$, which is assumed to be on resonance with the QD emission. In the frame rotating at frequency $\omega_X - D$, the system Hamiltonian is now
\begin{align}
	H_\Sys(t) = & -\hbar D \dyad{X} + \hbar g \qty(a^\dag \sigma + a \sigma^\dag )  + \\ \nonumber
	& + \qty{ \frac{\hbar}{2} \qty[ \Omega_1(t) e^{-i \delta_1 t} + \Omega_2(t) e^{-i \delta_2 t}] \sigma^\dag + \mathrm{h.c.} }
\end{align}
with $g$ the QD-cavity coupling strength.
We also add three Lindblad terms \cite{BreuerPetruccione} to the master equation, which account for photon leakage out of the cavity at a rate $\kappa$, spontaneous decay of the QD into non-cavity (background) modes at a rate $\Gamma_{\rm b}$, and pure dephasing at a rate $\gamma_{\rm d}$ induced by charge and nuclear spin fluctuations in the vicinity of the emitter. 
The master equation now reads
\begin{align}
	\dv{t} \rho_\Sys = 
	& - \frac{i}{\hbar} \comm{H_\Sys(t)}{\rho_\Sys} + \mathcal{K} \qty[\rho_\Sys] \nonumber \\
	& + \kappa \mathcal{L}_a[\rho_\Sys] + \Gamma_{\rm b} \mathcal{L}_{\sigma}[\rho_\Sys] + \gamma_{\rm d} \mathcal{L}_{\sigma^\dag \sigma}[\rho_\Sys]
\end{align}
with $\mathcal L_{A}[\rho] = A \rho A^\dag - \frac 1 2 \acomm{A^\dag A}{\rho}$.
We consider a micropillar device formed by sandwiching the QD between two stacks of DBR mirrors, whose performance has been optimized in previous work \cite{Wang2020_PRB_Biying, Wang2021_APL_Biying}.
Parameters for the microscopic modeling ($g$, $\kappa$, and $\Gamma_{\rm b}$) are extracted from optical simulations of the electromagnetic environment \cite{Wang2020_PRB_Biying}. We use $g = 0.041$\,THz, $\kappa = 0.46$\,THz, $\Gamma_{\rm b} = 0.45 \cdot 10^{-3}$\,THz, and the dephasing rate is set at $\gamma_{\rm d} = 0.13 \cdot 10^{-3}$\,THz.

The number $\mathcal{N}$ of single photons successfully reaching the collection optics is calculated as
\begin{equation}
	\mathcal N = \gamma_{\rm coll} \kappa \int_{t_0}^{+\infty} \dd{t} \expval{a^\dag(t) a(t)}
\end{equation}
where $\gamma_{\rm coll}$ is the fraction of photons emitted from the cavity that are successfully collected. We use $\gamma_{\rm coll} = 1$ for the two-color schemes and $\gamma_{\rm coll} = 0.5$ for resonant excitation, due to the need for cross-polarization filtering.
The indistinguishability of the emitted photons is determined as 
\begin{equation}
	\mathcal I = 1 - \frac
	{\int \dd{t} \int \dd{s} \qty[G^{(2)}_{\rm pop}(t, s) + g^{(2)}(t, s) - \qty|g^{(1)}(t, s)|^2]}
	{\int \dd{t} \int \dd{s} \qty[2 G^{(2)}_{\rm pop}(t, s) - \qty|\expval{a(t+s)} \expval{a^\dag(t)}|^2]}
\end{equation}
with $G^{(2)}_{\rm pop}(t, s) = \expval{a^\dag(t) a(t)} \expval{a^\dag(t+s) a(t+s)}$, $g^{(2)}(t, s) = \expval{a^\dag(t) a^\dag(t+s) a(t+s) a(t)}$, and $g^{(1)}(t, s) = \expval{a^\dag(t) a(t+s)}$ as detailed in Refs.\ \cite{Gustin2020, Gustin2018}.
For later convenience, we also calculate the single-photon purity $\mathcal P$ as \cite{Bracht2021, Gustin2020, Heinisch2023}
\begin{equation}
	\mathcal P = 1 - \frac
	{\int \dd{t} \int \dd{s} g^{(2)}(t, s)}
	{\int \dd{t} \int \dd{s} G^{(2)}_{\rm pop}(t, s)}
\end{equation}
which measures the probability of multi-photon emission.
Correlation functions are evaluated using the quantum regression theorem \cite{BreuerPetruccione}, which may overestimate the effect of phonon coupling \cite{Cosacchi2021, Fux2022}. One can thus interpret our result as a lower bound to $\mathcal I$.

Losses and decoherence may occur both during the \emph{exciton preparation} phase and the \emph{emission} phase. In the latter, some photons will be lost into background modes if $\Gamma_{\rm b} \ne 0$, resulting in a coupling efficiency to the cavity mode $\beta < 1$. 
This sets an upper bound $\mathcal N \leq \mathcal N^{\rm (UB)} = \beta$ for the case of pure single-photon emission \cite{footnote1}.
Similarly, $\mathcal I$ is deteriorated both by temporal indeterminacy in the excited state preparation via the time-jitter effect \cite{Kiraz2004}, and by phonon scattering and noise-induced dephasing during the emission phase.
To calculate the maximum performance that can be attained with such a cavity design, we artificially initialize the system in the state $\ket{\psi(t_0)} = \ket{X}$ as done in Ref.\ \cite{Iles-Smith2017}. We obtain $\mathcal N^{\rm (UB)} = 0.966$ and $\mathcal I^{\rm (UB)} = 0.975$.
These hard limits --- which can be understood in terms of the Franck-Condon factor as explained in Ref.\ \cite{Iles-Smith2017} --- are defined by dissipation and decoherence occurring \emph{after} excitation, and not by imperfections in the excitation.

\begin{figure}
	\centering
	\includegraphics[width=0.99\linewidth]{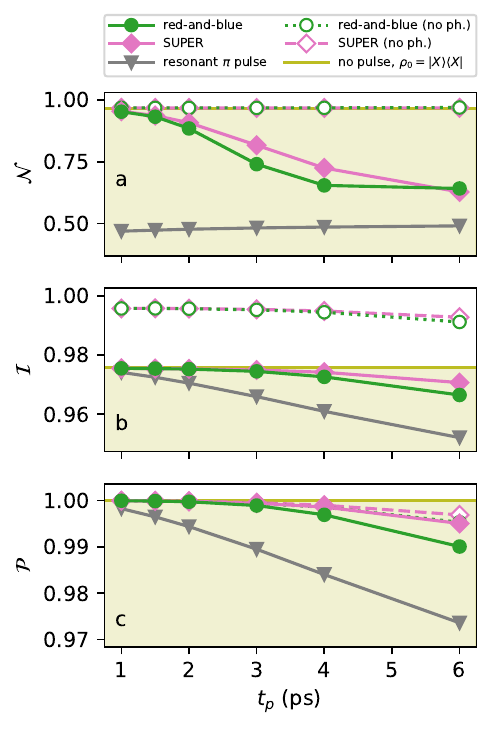}
	\caption{
		Figures of merit $\mathcal N$, $\mathcal I$, and $\mathcal P$ as a function of $t_p$ for a two-color and a resonant $\pi$ pulse of the same duration.
		The frequency detunings are determined as $\delta_1 = -\delta_2 = 6 / t_p$ for the red-and-blue dichromatic scheme, $\delta_1 = -6 / t_p$ and $\delta_2 = 3.49 \delta_1$ for the SUPER scheme, and $\delta=0$ for the resonant pulse.
		The upper bounds $\mathcal N^{\rm (UB)}$, $\mathcal I^{\rm (UB)}$ and $\mathcal P^{\rm (UB)}$ are also reported, and the regions below such bounds are marked with a shade.
		}
	\label{fig:7}
\end{figure}

In Figs.\ \ref{fig:7}a and \ref{fig:7}b, we show the values $\mathcal N^{\rm (UB)}$ and $\mathcal I^{\rm (UB)}$ calculated for an initially excited exciton with a yellow line.
The values $\mathcal N$ and $\mathcal I$ calculated starting from $\ket{\psi(t_0)} = \ket{G}$ and in the presence of the pumping laser will necessarily obey $\mathcal N \leq \mathcal N^{\rm (UB)}$ and $\mathcal I \leq \mathcal I^{\rm (UB)}$, and we highlight this accessible region with a yellow shade.
We then plot $\mathcal N$ and $\mathcal I$ for the red-and-blue and the SUPER two-color drives at constant $\eta$ and $R$ and different values of $t_p$, and a resonant $\pi$ pulse of the same duration.
For each data point, the two-color detunings $\delta_1$, $\delta_2$ are chosen as $\delta_1 = -\delta_2 = 6 / t_p$ for the red-and-blue dichromatic scheme, and $\delta_1 = -6 / t_p$ and $\delta_2 = 3.49 \delta_1$ for the SUPER scheme.

As expected, the best $\mathcal N$ is obtained for shorter pulses, where the performance is almost at the same level as the one calculated in the absence of phonons.
Interestingly, all data points for two-color excitation are well above the 0.5 limit for resonant excitation, which is set by the need for polarization filtering.
In particular, the value increases from $\mathcal N = 0.642$ (0.628) at $t_p=6$\,ps to $\mathcal N = 0.953$ (0.954) at $t_p=1$\,ps for the red-and-blue (SUPER) two-color excitation.
Importantly, the deviation $1 - \mathcal N$ is not a fundamental limitation of our scheme. Losses through spontaneous emission into background modes are calculated as $\mathcal N_{\rm b} = \Gamma_{\rm b} \int_{t_0}^{+\infty} \dd{t} \expval{\sigma^\dag(t) \sigma(t)}$,
which yields $\mathcal N_{\rm b} = 0.034$ at $t_p=1$\,ps for the two-color schemes.
An additional loss $1 - P_X$ is caused by imperfect exciton preparation as obtained previously for a bulk QD in the absence of any emission mechanism, and we indeed observe that $\mathcal N + \mathcal N_{\rm b} + (1 - P_X) = 1$. Note, however, that losses due to population inversion can be made arbitrarily small by resorting to shorter pulses, while background emission is unrelated to the pumping mechanism, and can be controlled using photonic engineering \cite{Gaal2022}.

Turning to the indistinguishability $\mathcal I$, we observe very high values $\mathcal I \geq 0.99$ for two-color schemes in the absence of phonon coupling (dashed curves in Fig.\ \ref{fig:7}b), with a slight decrease for longer pulses. The reason for this is attributed to possible re-excitation of the QD which becomes more likely for slow pulses, with a detrimental effect on two-photon interference visibility.
This is confirmed by the single-photon purity $\mathcal P$ shown in Fig.\ \ref{fig:7}c. Here, a value $\mathcal P < 1$ indicates the occurrence of multi-photon emission, while $\mathcal P^{\rm (UB)}=1$ is the upper bound corresponding to perfect anti-bunching, which is obtained with an initially excited emitter. 

As mentioned, when the effect of phonon coupling is taken into account (solid curves in Fig.\ \ref{fig:7}) the indistinguishability remains below the value calculated in the absence of phonons due to decoherence in the emission dynamics \cite{Iles-Smith2017}.
Nevertheless, we obtain an excellent performance for any choice of parameters, with $\mathcal I$ ranging from 0.966 at $t_p=6$\,ps up to 0.975 at $t_p=1$\,ps --- the latter corresponding exactly to the upper bound due to unavoidable phonon scattering during emission, as defined by the Franck-Condon factor \cite{Iles-Smith2017}.
It is noteworthy that the indistinguishability obtained under two-color schemes is better than the one under resonant excitation for any value of $t_p$, with all schemes converging towards $\mathcal I^{\rm (UB)}$ at short $t_p$.
This is again a consequence of the single-photon purity $\mathcal P$. As was recently explained in Ref.\ \cite{Heinisch2023}, two-color pumping is less likely to induce re-excitation of the QD than a resonant scheme due to its strongly off-resonant nature.
Indeed, in Fig.\ \ref{fig:7}c we always obtain $\mathcal P \geq 0.990$ using two-color schemes, with $\mathcal P \geq 0.999$ for $t_p \leq 3$\,ps. On the other hand, purity under resonant excitation ranges between $\mathcal P = 0.974$ (at $t_p=6$\,ps) and $\mathcal P = 0.998$ (at $t_p=1$\,ps).

\section{Conclusions}
\label{sec:conclusions}

In conclusion, we have discussed the role of phonon coupling on the performance of two-color pumping scheme such as the red-and-blue dichromatic excitation and the SUPER scheme.
We have shown that a necessary condition in order to exploit the full potential of two-color excitation is to enter the regime of phonon decoupling, where the exciton population oscillates on a time scale which is faster than the phonon relaxation time ad the pulse detuning are beyond the range of phonon spectral density.
Finally, our calculations demonstrate that two-color pumping schemes can push the performance of state-of-the-art SPSs towards $\mathcal N \cdot \mathcal I = 1$, in accordance with the requirements for scalable quantum technologies.
Our results establish two-color excitation as fundamental tool for SPS engineering, and illustrate the importance of pulse-shaping devices such as Spatial Light Modulators \cite{Karli2022}.

\begin{acknowledgments}

We thank Battulga Munkhbat for stimulating discussions, Martin A.\ Jacobsen for providing optical simulations of the micropillar SPS, and Doris Reiter for useful comments on the manuscript.
This work is funded by the European Research Council (ERC-CoG "UNITY", grant 865230) and by the Independent Research Fund Denmark (Grant No. DFF-9041-00046B).

\end{acknowledgments}

% \clearpage

\appendix

\section{Functional dependence of $P_X$}
\label{app:proof}

Here we show that, for the case with no phonon coupling ($\alpha=0$), the exciton population $P_X(t_1)$ after the laser pulse is a function of a limited set of parameters.
In the absence of phonons and any spontaneous decay, the dynamics is unitary and determined by $P_X(t_1) = \qty| \mel{X}{U(t_1, t_0)}{G} |^2$ with
\begin{equation}
	U(t_1, t_0) = \mathcal T \exp \qty[-\frac{i}{\hbar} \int_{t_0}^{t_1} \dd{u} H_\Sys(u)] .
\end{equation}
With a simple change of variable, this reads
\begin{widetext}
\begin{equation}
	U(t_1, t_0) = \mathcal T \exp \qty[- \frac{i}{2 \sqrt{\pi}} \int_{\frac{t_0}{t_p}}^{\frac{t_1}{t_p}} \dd{s} e^{-s^2} \qty( \Theta_1 e^{-i t_p \delta_1 s} + \Theta_2 e^{-i t_p \delta_2 s}) \sigma^\dag + \mathrm{h.c.} ]
\end{equation}
Since $e^{-s^2} \approx 0$ for large $\qty|s|$, one can safely extend the integration to $\pm \infty$ provided that $t_0$ and $t_1$ are chosen suitably --- in practice, it is sufficient to take $t_0/t_p \leq -3$ and $t_1/t_p \geq +3$. With the definitions $\eta = t_p \delta_1$ and $R = \delta_2 / \delta_1$ one has
\begin{equation}
	U(t_1, t_0) = \mathcal T \exp \qty[- \frac{i}{2 \sqrt{\pi}} \int_{-\infty}^{+\infty} \dd{s} e^{-s^2} \qty( \Theta_1 e^{-i \eta s} + \Theta_2 e^{-i R \eta s}) \sigma^\dag + \mathrm{h.c.} ]
\end{equation}
\end{widetext}
which indeed shows that $P_X(t_1) = f(\Theta_1, \Theta_2, \eta, R)$.
To obtain a symmetric red-and-blue configuration we set $R=-1$, thereby reducing the free parameters to three.

\section{Methods}
\label{app:methods}

\subsection{Weak coupling master equation}

We resort to the weak-coupling master equation to calculate the QD dynamics in the presence of phonon coupling \cite{Nazir2016}.
The master equation for the density operator $\rho(t)$ reads
\begin{equation}
	\dv{t} \rho(t) = - \frac{i}{\hbar} \comm{H_\Sys(t)}{\rho(t)} + \mathcal{K}(t) \qty[\rho(t)]
\end{equation}
where the phonon dissipation is given by
\begin{equation}
	\mathcal{K}(t) \qty[\rho(t)] = \int_0^{+\infty} \dd{s} C(s) \comm{\hat{X}(t-s,t) \rho(t)}{X} + \mathrm{h.c.}
\end{equation}
with $X = \sigma^\dag \sigma$, $\hat{X}(t-s,t) = U^\dag(t-s, t) X U(t-s, t)$ and 
\begin{equation}
	\label{eq:U_operator}
	U(t-s, t) = \mathcal T \exp \qty[-\frac{i}{\hbar} \int_{t}^{t-s} \dd{u} H_\Sys(u)]
\end{equation}
The environment correlation function reads
\begin{multline}
	C(s) = \int_0^{+\infty} \dd{\omega} J_{\rm ph}(\omega) \cdot \\
	\cdot \qty[\coth \qty(\frac{\hbar \omega}{2 \kappa_{\rm B} T})\cos(\omega s) - i \sin(\omega s)]
\end{multline}
with $T$ the temperature, which is set at $T=4$\,K throughout this work.

The dynamics is obtained by solving the master equation via a 4th-order Runge-Kutta algorithm, with initial condition $\rho(t_0) = \dyad{G}$. We set $t_0 = -3 t_p$ to make sure that $\Omega_j(t \leq t_0) \approx 0$.
A second Runge-Kutta algorithm is used to calculate $U(t-s, t)$ in Eq.\ \eqref{eq:U_operator} by solving
\begin{equation}
	\dv{s} U(t-s, t) = \frac{i}{\hbar} H_\Sys(t-s) U(t-s, t)
\end{equation}
at fixed $t$ as a function of $s$, with initial condition $\eval{U(t-s, t)}_{s=0} = I$ ($I$ being the identity). Finally, two-time correlation functions necessary for the indistinguishability calculation are evaluated using the quantum regression theorem \cite{BreuerPetruccione}.

\subsection{Polaron theory}

In the polaron theory, which is presented here for comparison, the Hamiltonian is diagonalized by applying the polaron transformation. This removes the QD-phonon interaction term [Eq.\ \eqref{eq:phonon_int}] by applying a unitary displacement to the phonon modes \cite{Nazir2016}.
The system dynamics is then obtained by solving the polaron master equation,
\begin{equation}
	\label{eq:polaron_ME}
	\dv{t} \rho(t) = - \frac{i}{\hbar} \comm{H_\Sys(t)}{\rho(t)} + \mathcal{K}_{\rm pol}(t) \qty[\rho(t)]
\end{equation}
The system Hamiltonian, in a frame rotating at frequency $\omega_X - D$, is
\begin{equation}
	\label{eq:polaron_H}
	H_\Sys(t) = \frac{\hbar}{2} B \sum_{j=1,2} \qty[ \Omega_j(t) e^{-i \delta_j t} \sigma^\dag + \Omega_j(t) e^{+i \delta_j t} \sigma] ,
\end{equation}
Here the quantity $B = \exp[-\frac 1 2 \phi(0)]$ is a phonon-induced re-normalization of the laser pulse amplitude, where $\phi(s)$ is the phonon correlation function
\begin{multline}
	\phi(s) = \int_0^{+\infty} \dd{\omega} \frac{J_{\rm ph}(\omega)}{\omega^2} \cdot \\
	\cdot \qty[\coth \qty(\frac{\hbar \omega}{2 \kappa_{\rm B} T})\cos(\omega s) - i \sin(\omega s)]
\end{multline}

Note that the polaron transformation automatically produces the frequency shift $\omega_X \to \omega_X - D$, which is thus not explicitly included in Eq.\ \eqref{eq:polaron_H}.

The second term in Eq.\ \eqref{eq:polaron_ME} is the polaron dissipator,
\begin{multline}
	\mathcal{K}_{\rm pol} \qty[\rho(t)] = \frac{1}{\hbar^2} \int_0^{+\infty} \dd{s} \cdot \\
	\cdot \sum_{i=x,y} C_{ii}(s) \comm{\hat{A_i}(t-s,t) \rho(t)}{A_i(t)} + \mathrm{h.c.}
\end{multline}
Here, the time dependent operators $A_i(t)$ are given by
\begin{align}
	A_x(t) & = \frac{\hbar}{2} B \sum_{j=1,2} \qty[ \Omega_j(t) e^{-i \delta_j t} \sigma^\dag + \Omega_j(t) e^{+i \delta_j t} \sigma] , \\
	A_y(t) & = i \frac{\hbar}{2} B \sum_{j=1,2} \qty[ \Omega_j(t) e^{-i \delta_j t} \sigma^\dag - \Omega_j(t) e^{+i \delta_j t} \sigma]
\end{align}
and $\hat{A_i}(t-s,t) = U^\dag(t-s, t) A_i(t-s) U(t-s, t)$, with $U(t-s, t)$ as in Eq.\ \eqref{eq:U_operator}.
The environment correlation functions are
\begin{align}
	C_{xx}(s) & = B^2 \qty{ \cosh[\phi(s)] - 1} , \\
	C_{yy}(s) & = B^2 \sinh[\phi(s)]
\end{align}
Again, we use a 4th-order Runge-Kutta algorithm to solve the polaron master equation, with initial condition $\rho(t_0 = -3 t_p) = \dyad{G}$.

\section{Effect of the temperature}
\label{app:temperature}

In this section, we show additional evidence for the phonon decoupling effect by analyzing the exciton population $P_X$ after the red-and-blue dichromatic pulse as a function of the temperature.
We consider three different configurations, all of them with constant $t_p \delta = 6$:
\begin{itemize}
	\item \textbf{slow pulse} (red star in Fig.\ 2a of the main text):\\ $(t_p, \delta)$ = (6\,ps, 1\,THz), \\ $(\Theta_b, \Theta_r) = (2.12\pi, 6.96\pi)$
	\item \textbf{fast pulse} (purple star in Fig.\ 2d of the main text):\\ $(t_p, \delta)$ = (1\,ps, 6\,THz), \\ $(\Theta_b, \Theta_r) = (1.80\pi, 6.96\pi)$ 
	\item \textbf{ultra-fast pulse} (not shown in the main text):\\ $(t_p, \delta)$ = (0.2\,ps, 30\,THz), \\ $(\Theta_b, \Theta_r) = (1.80\pi, 6.96\pi)$
\end{itemize}

\begin{figure}
	\centering
	\includegraphics[width=0.99\linewidth]{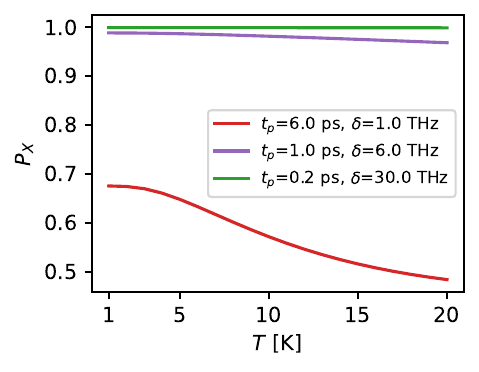}
	\caption{
		Exciton population after a slow (red), fast (purple), and ultra-fast (green) pulse as a function of the temperature. See text for details.
	}
	\label{fig:8}
\end{figure}

As reported in Fig.\ \ref{fig:8}, the performance of the slow pulse falls significantly with increasing temperature, as expected (from 0.675 at $T=1$\,K to 0.484 at $T=20$\,K). This is due to phonon scattering becoming more and more relevant at higher temperature, which results in a smaller population inversion.

On the other hand, temperature has little influence on the fast pulse, with a mere 2\% decrease in $P_X$ between $T=1$\,K and $T=20$\,K (from 0.988 to 0.968). We attribute this fact to the decoupling mechanism, which quenches the detrimental effect of phonons even at relatively high temperature.

An ultra-fast pulse with $t_p$ goes even deeper into the decoupling regime, as shown by the green curve in Fig.\ \ref{fig:8}. Here, $P_X$ stays constant at 0.999 for the entire range $T \in$ (1\,K, 20\,K).
This last configuration is unpractical for applications, but is presented here to demonstrate the phonon decoupling.

\section{Comparison of different methods}
\label{app:comparison}

\begin{figure}
	\centering
	\includegraphics[width=0.99\linewidth]{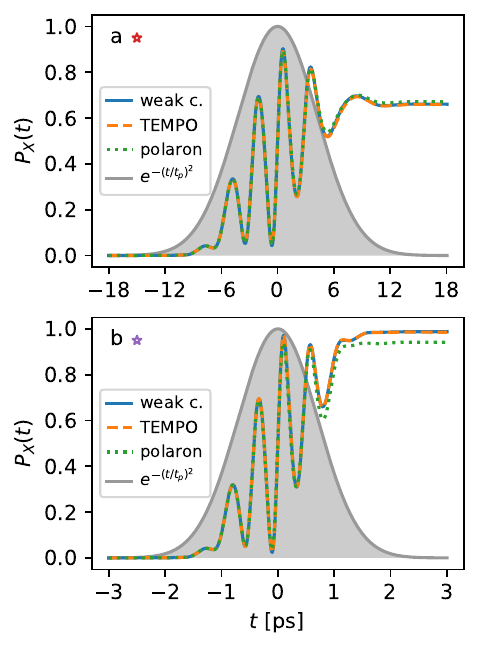}
	\caption{
		Comparison of the time evolution $P_X(t)$ predicted by the weak-coupling master equation, the polaron master equation, and the TEMPO algorithm after a slow (panel a) and fast (panel b) pulse. See text for details, and colored markers in Fig.\ \ref{fig:3}a and \ref{fig:3}d of the main text.
	}
	\label{fig:9}
\end{figure}

Predictions of the weak-coupling theory are compared here with results obtained from the polaron master equation and the more advanced TEMPO method \cite{Strathearn2018, OQuPy}.
The latter offers the benefit of being numerically exact at the cost of a larger numerical burden and coding complexity.
As such, it is a valuable tool to explore new physics in a non Markovian regime, or to assess the performance of approximate methods such as the weak-coupling and polaron theories.
Here, TEMPO calculations are performed using the open source Python package \textsc{OQuPy} \cite{OQuPy}.

Figures \ref{fig:9}a and \ref{fig:9}b show the excited state population $P_X(t)$ predicted by the three methods after the slow and fast red-and-blue pulse respectively (see previous section).
The weak-coupling prediction is in perfect quantitative agreement with the TEMPO calculation, both for slower and faster driving. This certifies that the weak-coupling master equation is indeed sufficient to accurately describe the pumping dynamics even in the presence of fast oscillations, provided that the phonon coupling is not too strong and the temperature is sufficiently low.
On the other hand, we observe a deviation of the polaron results from the other two lines. For the case of fast driving ($t_p = 1$\,ps, Fig.\ \ref{fig:9}b), we obtain $P_X^{\rm (pol)} = 0.941$, in contrast with $P_X = 0.987$ obtained from the weak-coupling and TEMPO calculations.
Thus, while being in qualitative agreement with the other two methods, the polaron underestimates the final probability by a factor $\sim 5\%$, which is a significant difference in the quest for a QD excitation scheme with near-unity efficiency.
The reason is that the polaron theory overestimates phonon-induced dissipation when the exciton population oscillates too fast, i.e.\ on a time scale that is shorter than the phonon relaxation time --- a very relevant case in this work.

\bibliography{biblio.bib}

\begin{thebibliography}{51}
\bibitem [{foo()}]{footnote1}%
  \BibitemOpen
  \href@noop {} {}\bibinfo {note} {$\mathcal N$ can actually be larger than
  $\beta$ if multi-photon emission occurs}\BibitemShut {NoStop}%
\end{thebibliography}

\end{document}